\documentclass[final,journal,letterpaper]{IEEEtran4PSCC}

 \usepackage[hidelinks]{hyperref}
 \usepackage{varioref} 
 \usepackage{wrapfig} 
 \usepackage{threeparttable} 
 \usepackage{dcolumn} 
  \newcolumntype{d}{D{.}{.}{-1}}
 \usepackage{nomencl} 
  \makeglossary
 \usepackage{subfigmat} 
 \usepackage{fancyvrb} 
  \fvset{fontsize=\footnotesize,xleftmargin=2em}
 \usepackage{lettrine} 
 \usepackage{algorithm,algorithmic}

 \usepackage{epsfig}
 \usepackage{epstopdf}
 \usepackage{subfigure}
 \usepackage{latexsym}
 \usepackage{amsmath}
 \usepackage{amsfonts}
 \usepackage{amssymb}
 \usepackage{mathrsfs}
 \usepackage{cases}
 \usepackage{array}
 \usepackage{color}
 \usepackage{soul}
 \setulcolor{red}
 \sethlcolor{yellow}
 \newcommand{\comment}[1]{}

 \usepackage{url}
 \usepackage[noadjust]{cite}
 \usepackage{graphicx}
 \usepackage{psfrag}
 \usepackage{color}
   \usepackage{multirow}
 \usepackage{ragged2e}
 \usepackage{graphicx}

\makeatletter
\let\old@ps@headings\ps@headings
\let\old@ps@IEEEtitlepagestyle\ps@IEEEtitlepagestyle
\def\psccfooter#1{%
    \def\ps@headings{%
        \old@ps@headings%
        \def\@oddfoot{\strut\hfill#1\hfill\strut}%
        \def\@evenfoot{\strut\hfill#1\hfill\strut}%
    }%
    \def\ps@IEEEtitlepagestyle{%
        \old@ps@IEEEtitlepagestyle%
        \def\@oddfoot{\strut\hfill#1\hfill\strut}%
        \def\@evenfoot{\strut\hfill#1\hfill\strut}%
    }%
    \ps@headings%
}
\makeatother

\psccfooter{%
        \parbox{\textwidth}{\hrulefill \\ \small{23rd Power Systems Computation Conference} \hfill \begin{minipage}{0.2\textwidth}\centering \vspace*{4pt} \includegraphics[scale=0.06]{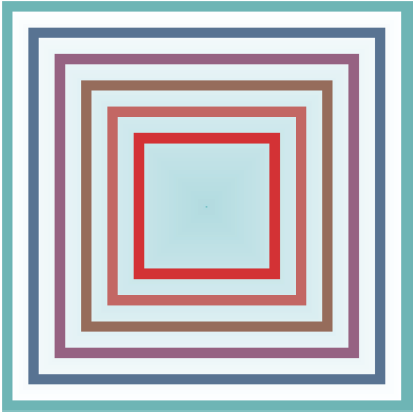}\\\small{PSCC 2024} \end{minipage} \hfill \small{Paris, France --- June 4 -- 7, 2024}}%
}

\begin{document}
%
\title{Toward Intelligent Emergency Control for Large-scale Power Systems: Convergence of Learning, Physics, Computing and
Control}

\author{\IEEEauthorblockN{Qiuhua Huang\IEEEauthorrefmark{1}\IEEEauthorrefmark{2},
Renke Huang\IEEEauthorrefmark{2},
Tianzhixi Yin\IEEEauthorrefmark{2},
Sohom Datta\IEEEauthorrefmark{2},
Xueqing Sun\IEEEauthorrefmark{2},
Jason Hou\IEEEauthorrefmark{2},
Jie Tan\IEEEauthorrefmark{3},
Wenhao Yu\IEEEauthorrefmark{3},
Yuan Liu\IEEEauthorrefmark{2},
Xinya Li\IEEEauthorrefmark{2},
Bruce Palmer\IEEEauthorrefmark{2}, 
Ang Li\IEEEauthorrefmark{2}, 
Xinda Ke\IEEEauthorrefmark{2}, 
Marianna Vaiman\IEEEauthorrefmark{3},
Song Wang\IEEEauthorrefmark{5},
Yousu Chen\IEEEauthorrefmark{2}\\}
\IEEEauthorblockA{\IEEEauthorrefmark{1} Colorado School of Mines, Golden, CO, USA, qiuhuahuang@mines.edu\\}
\IEEEauthorblockA{\IEEEauthorrefmark{2} Pacific Northwest National Laboratory, Richland, WA, USA, yousu.chen@pnnl.gov\\}
\IEEEauthorblockA{\IEEEauthorrefmark{3} Google DeepMind,
Mountainview, CA, USA, jietan@google.com\\}
\IEEEauthorblockA{\IEEEauthorrefmark{4}V\&R Energy System Inc, Los Angles, CA, USA, marvaiman@vrenergy.com\\}
\IEEEauthorblockA{\IEEEauthorrefmark{5}Portland General Electric,Portland, OR, USA, song.wang@pgn.com}
}

\maketitle

\begin{abstract}

This paper has delved into the pressing need for intelligent emergency control in large-scale power systems, which are experiencing significant transformations and are operating closer to their limits with more uncertainties.  Learning-based control methods are promising and have shown effectiveness for intelligent power system control. However, when they are applied to large-scale power systems, there are multifaceted challenges such as scalability, adaptiveness, and security posed by the complex power system landscape, which demand comprehensive solutions. The paper first proposes and instantiates a convergence framework for integrating power systems physics, machine learning, advanced computing, and grid control to realize intelligent grid control at a large scale. Our developed methods and platform based on the convergence framework have been applied to a large (more than 3000 buses) Texas power system, and tested with 56000 scenarios. Our work achieved a 26\% reduction in load shedding on average and outperformed existing rule-based control in 99.7\% of the test scenarios. The results demonstrated the potential of the proposed convergence framework and DRL-based intelligent control for the future grid.



\end{abstract}


\thanksto{\noindent The Pacific Northwest National Laboratory (PNNL) is operated by Battelle for the U.S. Department of Energy (DOE) under Contract DE-AC05-76RL01830. This work was supported by DOE ARPA-E OPEN 2018 Program.
\newline Qiuhua Huang, Renke Huang, Xueqing Sun, and Xinya Li were with PNNL while performing the research work for this paper.
\newline
$\ast$ \textit{Corresponding authors: Qiuhua Huang, Renke Huang}}

\section{Introduction}\label{sec:intro}

As power systems are undergoing a significant transformation with more uncertainties, less inertia, closer to operation limits, and more frequent extreme weather events such as hurricanes and heat waves, there is an increasing risk of large outages. Existing emergency controls are either armed by human operators in the control rooms or triggered by rule-based settings. Consequently, they suffer from either slow response due to human decision-making or ineffectiveness in changing operating conditions. Thus, there is an imperative need to enhance the intelligence (i.e., mainly automation and adaptiveness) of grid emergency control to maintain system security and stability. 

Some notable progress has been made in developing Artificial Intelligence (AI)-based or data-driven power system emergency control solutions in recent years. Deep reinforcement learning (DRL) represents one of the latest developments in AI for sequential decision-making and has been utilized or developed for grid stability and emergency control applications\cite{Yan_Frequency_DRL,huang2019loadshedding_DRL,chen2020model}. Recent review papers \cite{glavic2019DRL4GC, Chen2022RL,li2023DRL} show that deep reinforcement learning (DRL) methods have been developed and applied for various grid controls, providing fast and effective solutions in power system stability and emergency control applications and overcome some limitations with traditional rule-based and optimization-based methods. Readers are referred to them for a general introduction to DRL and DRL applications in power systems. There are \textit{multifaceted challenges such as scalability, adaptiveness, and security} posed by the complex power system landscape to achieve DRL-based intelligent wide-area or system-level grid emergency control. Existing works mostly focused on addressing one of these challenges. To tackle such grand challenges, we believe that single-domain advancements are inadequate because solutions designed for solving one challenge may not be compatible with solutions for another one. Furthermore, the scalability challenge has not been properly solved, and existing works are mostly developed for and/or tested with small transmission or distribution systems and only a few scenarios. For example, the largest system for DRL-based stability control prior to this work is the IEEE 300-bus test system with limited operation scenarios being considered\cite{huang2022accelerated}. To the best of our knowledge, none of them tackles critical issues such as scalability, adaptability, and security comprehensively, and is tested with realistic, large-scale power systems with diverse operation conditions.  This means a large gap between existing research efforts and real-world applications in this area. 

We need multi-domain advancements to address the gap, thus our key idea focuses on a framework for achieving \textbf{convergence} of the state-of-the-art reinforcement learning, physics of power systems, advanced computing and grid control methods, which is illustrated in Fig. \ref{fig:convergency} and has not been reported in the literature. We instantiate the convergence framework by developing a suite of comprehensive methods and tools, as summarized in Fig. \ref{fig:key_methods}. Our final solutions are comprehensively tested on a synthetic Texas power system with more than 3000 buses. We considered 14000 scenarios for training and tested the trained control policy with 56000 scenarios that are unseen in the training dataset, and compared with the existing rule-based under-voltage load-shedding (UVLS) control scheme. The results show that our approach helps reduce the load shedding by 26\% on average, and is better than the baseline in 99.7\% of the test scenarios in terms of the control objective value (i.e., total reward). This demonstrates great progress toward intelligent emergency control for large-scale power systems. 

The main contributions of this paper include:
\begin{enumerate}
    \item proposing and instantiating a convergence framework for integrating DRL, power systems physics,  advanced computing and grid control to achieve intelligent grid control at a large scale. 
    \item development of a scalable and high-performance computing (HPC)-compatible grid simulation environment for training and testing large-scale learning-based control solutions and it is open-sourced.
    \item a three-stage DRL training method including a two-stage curriculum learning method for distributed training of DRL-based control policies for each zone, followed by one-stage coordinated training of all control policies. 
    \item unprecedented large-scale training, testing of DRL-based grid emergency control with a large (more than 3000 buses) synthetic Texas system, demonstrating the potential of proposed convergence framework and DRL-based intelligent control for future grids.
\end{enumerate}

The rest of the paper is organized as follows: Section \ref{sec:problem} discusses the problem and our proposed framework and key methods. 
Section \ref{sec:implementation} provides more details of implementation of the convergence framework. 
Section \ref{sec:results} presents the training and testing results. 
The paper is concluded in Section V.

\section{Problem statement and our proposed convergence framework}\label{sec:problem}

As discussed in the introduction section, we have to address multifaceted challenges such as scalability, adaptiveness, and security to successfully develop intelligent grid stability and emergency control.  Different from previous work focusing on addressing a single challenge without considering the compatibility and synergy of the underlining methods and technologies, we highlight \textit{systematic solutions} by making technological advancements in the following key areas in a cohesive manner. 
\begin{enumerate}
    \item \textbf{Learning}: AI and machine learning can provide some unique learning from (big) data and decision-making capabilities that are essential for real-time monitoring, analysis, and control of modern power systems. Successful AI-based control applications require data from the physical world, existing control design principles, and computational resources and platforms.
    \item \textbf{Computing}: Advanced computational algorithms and tools are essential for power system analysis, control design and verification, and supporting various AI applications in power systems.
    \item \textbf{Control}: Grid control strategies are critical for maintaining grid stability and security. However, most existing controls are designed based on physics-based models for a few deterministic operation patterns and do not operate well under vastly changing operation conditions. AI-based methods can complement existing physics-based control designs.
    \item \textbf{Physics}: Power system physics provides the essential models, data, and domain knowledge for the areas above. Advanced computational methods and AI tools help engineers gain insightful information from big data in power systems. Physics-informed machine learning is an example of combining the best of both worlds.
\end{enumerate}

\begin{figure}[!t]
  \centering
\includegraphics[width=0.40\textwidth]{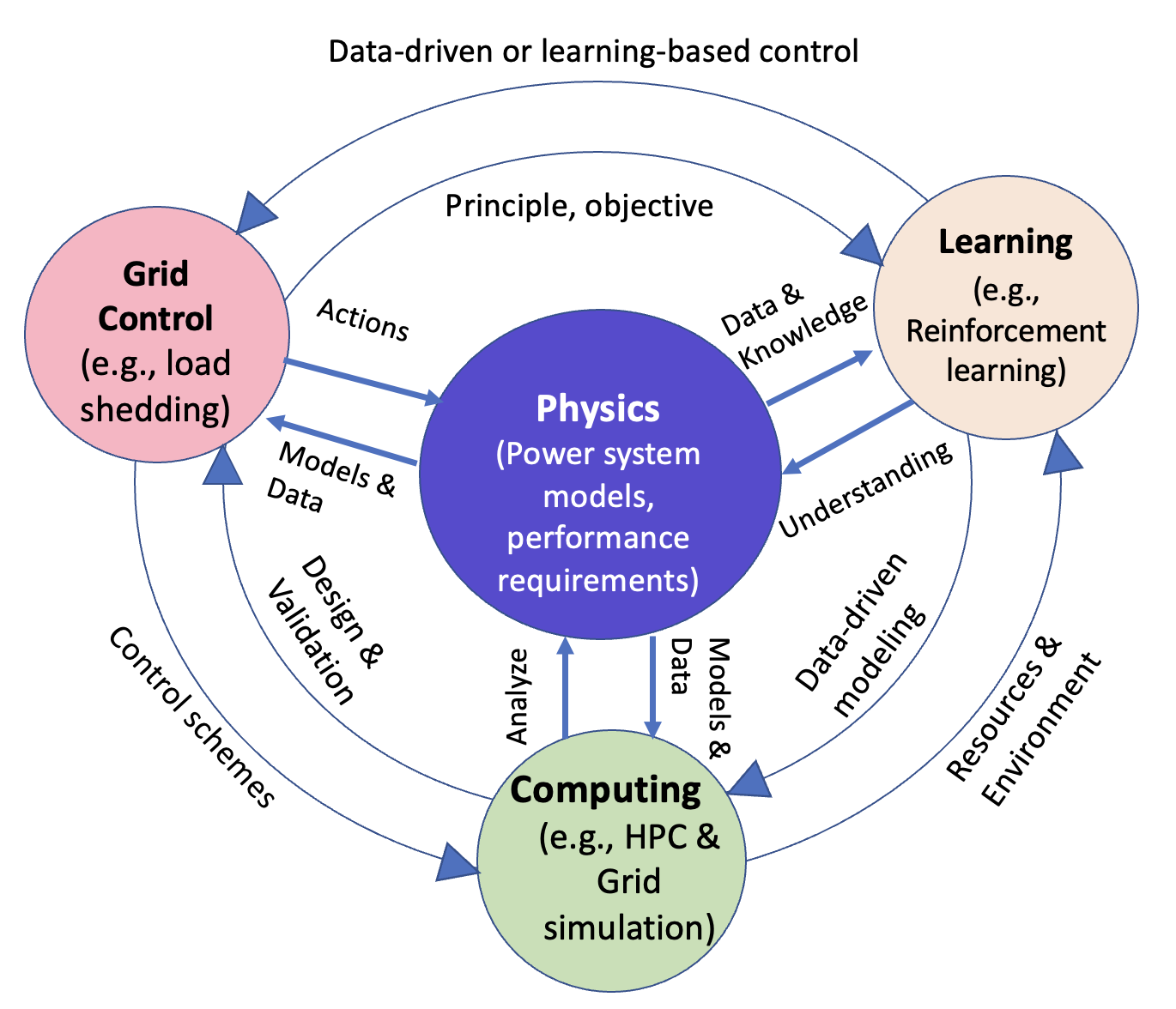}
  \caption{Convergence of power system physics, advanced computing, machine learning and grid control for achieving intelligent control for large-scale power systems}
  \label{fig:convergency}
 \end{figure}

It can be seen that there is a strong synergy among these technologies (also see Fig. \ref{fig:convergency}). However, to the best of our knowledge, few effort has been focused on the convergence of them to realize their full potential for intelligent emergency control for large-scale power systems. Therefore, we first develop a convergence framework by considering their inherent synergy in the context of power system stability and emergency control. The framework is shown in Fig. \ref{fig:convergency}. In the following sections, we will discuss the details of instantiating and implementing the framework.

\section{Implementation of the convergence framework}{\label{sec:implementation}}

\begin{figure*}[!t]
\centering
\includegraphics[width=15cm]{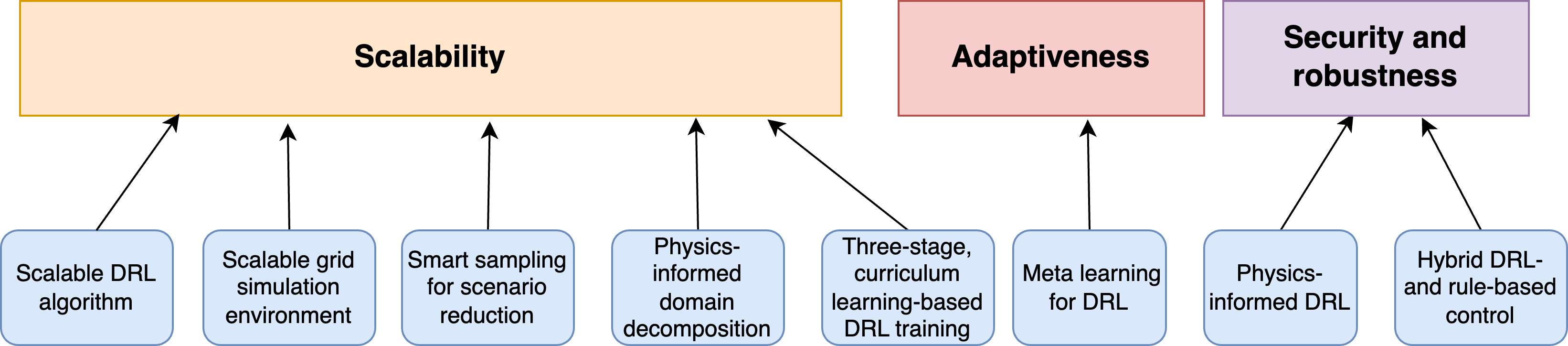}
\caption{Key methods based on the the convergence of physics, learning, computing and control}
\label{fig:key_methods}
\end{figure*}

With the convergence framework above as guidance, we instantiate it by developing several interdisciplinary methods to tackle the scalability, adaptability, and security challenges, as shown in Fig. \ref{fig:key_methods}. Below is a brief summary, with more discussions of the development of these key methods in the following subsections.

First, several key factors including computational complexity of large-scale power system dynamic simulation, high dimensional action space, large number of power system operation conditions and fault scenarios all contribute to the scalability challenge, which makes it significantly difficult to address. Therefore, we developed several complementary methods ranging from an HPC-compatible, highly scalable grid simulation environment, and a scalable DRL algorithm, to scenario reduction to overcome it.

Second, since power systems are always changing, there are a large number of grid operation conditions. We developed smart sampling techniques to identify representative scenarios for training the DRL model. In addition, we incorporated the meta-learning method into DRL to ensure it can learn from past experiences and quickly adapt to new scenarios. 

Third, like other AI methods, DRL-based methods still do not have a strict performance guarantee for some corner cases. We developed a hybrid DRL-and rule-based control method to ensure the final solution can perform better than existing rule-based control solutions most of the time, and have a fallback control scheme for other extreme cases. 

\subsection{High-performance grid simulation platform and DRL algorithm}\label{sec:hpc platform}

\begin{figure}[!t]
  \centering
\includegraphics[width=0.35\textwidth]{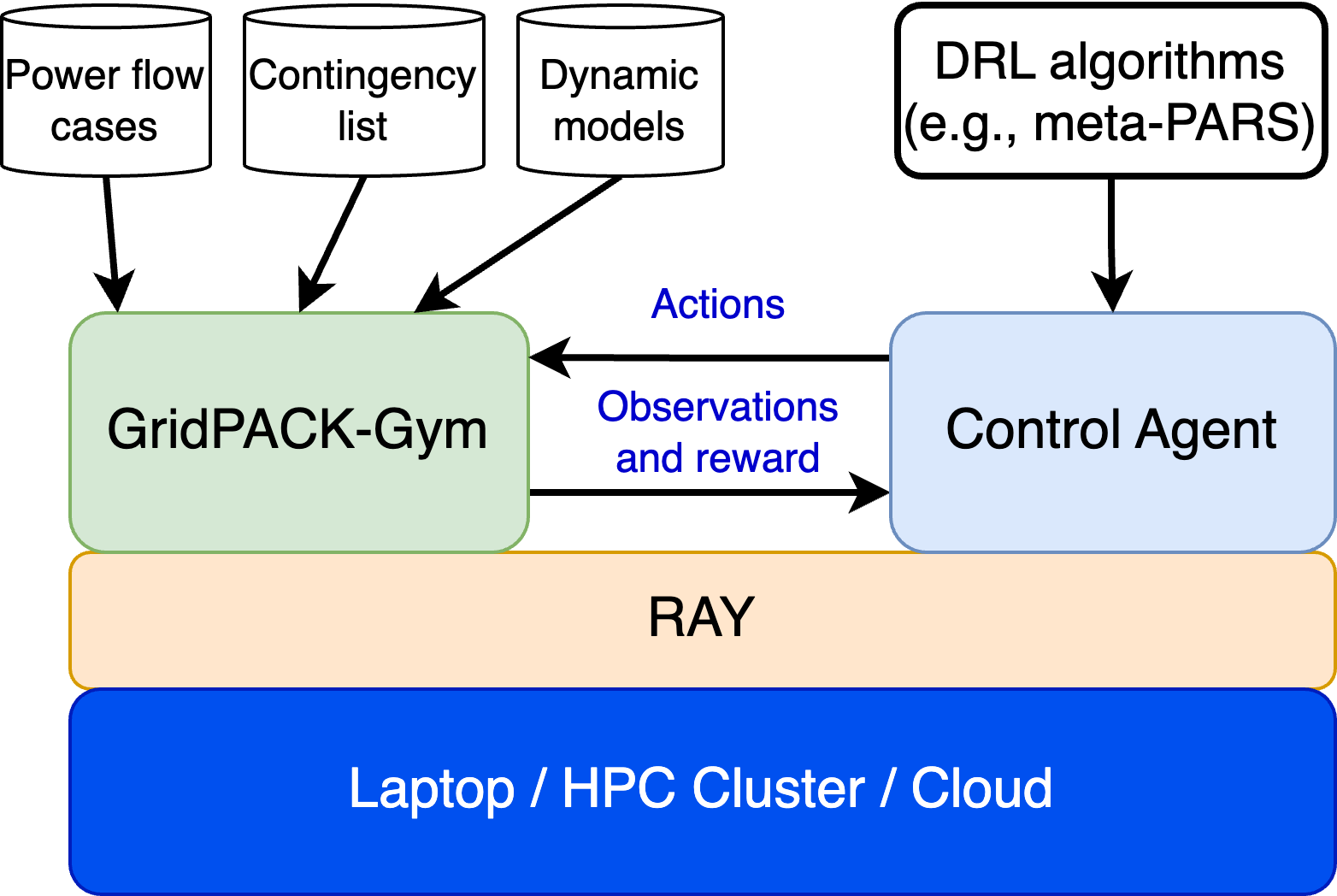}
  \caption{A high-performance grid simulation and reinforcement learning platform}
  \label{fig:hadrec_platform}
 \end{figure}

As shown in Fig. \ref{fig:hadrec_platform}, DRL-based control policies are primarily first trained in a simulated environment instead of in the real-world power grid due to security concerns. Dynamic simulation of complex power systems requires solving large-scale differential algebraic equation models, which is computationally intensive. As such, for large-scale RL training and testing, a common bottleneck is environment execution, which is often the slowest part of the whole system \cite{weng2022envpool}. Existing RL environments such as RLGC\cite{huang2019loadshedding_DRL} for grid control are designed for small-scale research and demonstration purposes.

To address this performance bottleneck, we developed the first high-performance simulation and learning platform for grid stability control. The platform architecture is shown in Fig. \ref{fig:hadrec_platform}. First of all, the platform is developed based on Ray\cite{Ray}, which is an open-sourced compute framework for scaling AI and Python workloads and supports various computing infrastructures from laptops to high-performance computing (HPC) clusters to clouds. This facilitates DRL-based control development from fast prototyping and testing to large-scale applications with the same Python codes. Secondly, we develop \textbf{GriPACK-Gym}(shown at the top of Fig. \ref{fig:hadrec_platform}), a scalable RL environment based on the high-performance grid simulator GridPACK \cite{GridPACK}. We developed new functions for setting up a grid environment for DRL training and an extra layer of Application Programming Interface (APIs) on top of GridPack APIs to make it fully compatible with the OpenAI Gym interface. GridPACK-gym can scale up to thousands of computing cores, which is the first of its kind. It has been open-sourced along with GridPACK on GitHub\cite{GridPACK-source}. 

To fully leverage available computing resources and the capabilities of GridPACK-Gym, we developed a highly scalable DRL algorithm called PARS in our earlier work\cite{huang2022accelerated} that can easily scale to thousands of computing cores. In light of the growing uncertainties and operation envelopes in power systems, we further enhance the adaptiveness of the PARS algorithm by integrating the meta-learning method into it and create a novel deep meta-reinforcement learning algorithm  \cite{huang2022DMRL} that is named \textbf{meta-PARS} in this paper. The incorporation of meta-learning enables the DRL-based controller (agent) to quickly learn or update a low-dimensional latent representation of the complex grid operation context to adapt the control strategy to new grid operation scenarios. We have made these novel DRL algorithms open-sourced along with the training codes on GitHub in \cite{HADREC-source}.

These are combined with other techniques introduced later in this section to realize successful training of the DRL-based control agent for intelligent grid emergency control. 

\subsection{Incorporate physics into DRL agent and training process}

There are significant amount of physical knowledge and best practices developed in the power system community for analyzing and operating the grid. On the other hand, recent AI research showed that it requires a massive amount of data for AI agents to learn even fundamental physical concepts or laws. Thus, a rising AI research direction is physics-informed machine learning which aims to leverage physics to jump-start the training of machine learning models and make it more robust. It is highly desirable to incorporate power system physics into DRL agent models and training processes.

First, we noticed that power system performance requirements such as voltage stability criteria \cite{WECC_voltage} are good prior knowledge that can be utilized to regulate the behavior of the neural network-based control policy. For the voltage stability emergency control application that we will consider as an example in our test cases, we incorporated the system transient voltage stability and recovery performance requirements into the DRL agent model as an action mask\cite{du2022PIES}. It can help block or filter out unfavorable or unnecessary control actions, thereby enhancing both the DRL training efficiency and robustness of the DRL-based control policy. More details can be found in our recent work \cite{du2022PIES}.

\begin{figure}[!t]
  \centering
\includegraphics[width=0.30\textwidth]{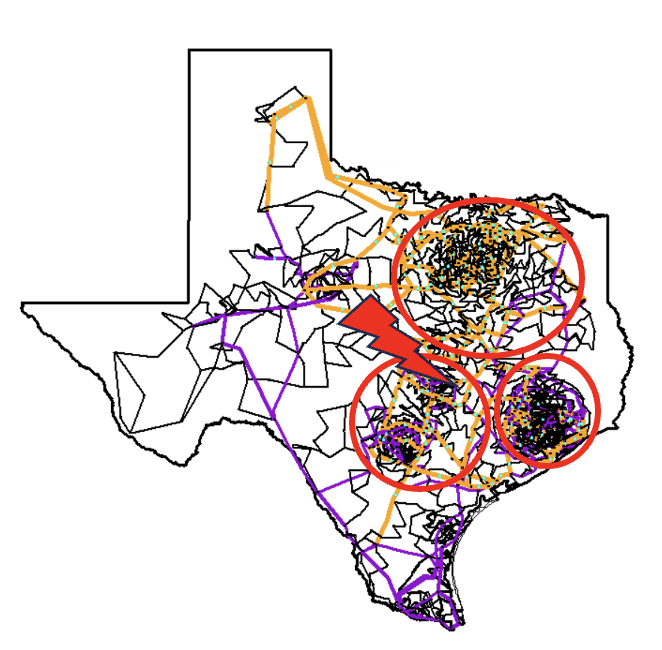}
  \caption{Physics-informed spatial decomposition for training (Figure adapted from \cite{xu2017creation})}
  \label{fig:decomposition}
 \end{figure}

Second, domain decomposition is an effective approach to reducing problem complexity and allows tackling each subdomain problem individually in a more tangible way. It should also be noted that many power system stability problems have localized impacts or can be solved by controlling generators or loads near the fault that leads to stability problems. In light of both, we proposed and developed a physics-informed spatial decomposition method in our DRL training process. The basic idea is illustrated in Fig. \ref{fig:decomposition} with the synthetic Texas power grid as the background. There are three main load centers in Texas that are vulnerable to fault-induced voltage stability issues. We first train individual DRL-based controllers for three load centers, which helps reduce the size of the action and observation spaces and makes the DRL-based controllers easier to train. However, as shown in Fig. \ref{fig:decomposition}, faults near the boundary between two load centers could result in voltage stability problems in two or more load centers and require coordinated control actions among them. Therefore, we need to coordinate these individually trained controllers, which will be addressed in the next subsection.

\subsection{A three-stage, curriculum learning-based RL method for overcoming difficult exploration problems}

RL agents generally learn by the so-called ``exploration and exploitation'' technique.  For RL training at the scale of hundreds of dimensional action space and observation space, tens of thousands of scenarios even after using down-sampling techniques, it is common to have difficult exploration problems where there are some difficult-to-train cases. There is no exception to our problem.

Along with the proposed physics-informed spatial decomposition method for training, we developed a three-stage DRL training method for overcoming the abovementioned problem and achieving successful training on a large-scale power system with more than 3000 buses and over 50 GW generation and load. First, we leverage the physics-informed spatial decomposition discussed before to create and train control policies for each control zone individually. 

Even with such a decomposition,  we noticed the training can still be very difficult in some cases in each zone. Therefore, we adapt the curriculum learning concept\cite{wang2021curriculum} to make the training more smooth (by focusing on solving easy cases first before solving difficult cases). In the conventional setup of curriculum learning, the training starts with a subset of ``easy'' cases and gradually switches to ``difficult'' cases. However, we found it difficult to pre-define the ``easy'' and ``difficult'' categories at the beginning across very different power flow conditions and fault scenarios. Therefore, we developed a two-stage curriculum learning method. In the first stage, all power flow cases and fault scenarios are treated equally and randomly sampled. The goal for this stage is twofold: 1) training a control policy that is good enough for a majority of the cases; 2) identifying the difficult-to-train cases where the control policy failed to achieve good performance. Then, in the second stage, difficult-to-train cases will be determined dynamically by the evaluation part of the proposed PARS algorithm if the total reward value is less than a threshold.  A  fixed number of difficult-to-train cases will be sampled and combined with other cases to form the full training set for the second stage. 

At the final stage, we perform coordinated training to combine them together in one training to enable them to learn to consider the mutual impacts of the control actions for each zone and hence coordinate their control actions.   

\subsection{Smart sampling for scenario reduction}{\label{ssec:sampling}}

In this paper, each grid scenario for training and testing DRL-based controllers is a combination of one power flow case for representing the grid operation condition, one contingency (defined jointly by fault location and fault duration) and one system dynamic model. We consider one system dynamic model for all studies, which is consistent with the current industry planning and operation practice. For large-scale power systems with high penetration of renewable energy resources, there are a large number of potential grid operation conditions in the form of power system snapshots. Thus, to deal with the scenario scalability challenge, it is necessary to down-sample a relatively small but representative subset of power flow cases. However, there are challenges in handling high-dimensional data with hierarchical structures and challenges arising from the non-parametric solution space where configurations are complicated by non-Gaussian behaviors and nonlinear cross-dependence in complex dynamic systems like power grids. We developed a hierarchical Latin Hypercube Sampling (LHS) technique for smart sampling, considering free-form distributions of system load, generator commitment status, and generation levels\cite{sun2021sampling}. The method can represent the probability distributions of the original high-dimensional data using fewer samples while respecting the hierarchical data-dependency among the variables. The approach is demonstrated using a large-scale Texas system and is used for scenario reduction in Section \ref{sec:results}.

For selecting a representative contingency list, it is important to have good coverage across the system and include both normal and severe contingencies to represent the nature and distribution of contingencies in the real-world system. We adopt the critical clearing time (CCT) as a metric for ranking the contingency locations and perform system-wide CCT analysis for all high-voltage buses. Then we down-sample the contingency fault locations by considering both the CCT and location in the system. Fault duration is sampled from the range between normal fault clearing time (3-4 cycles) to delayed fault clearing time (up to 25 cycles).

\subsection{DRL- and rule-based hybrid grid emergency control}

Electric utilities and power system operators have widely used rule-based control methods such as under-voltage load shedding (UVLS) or under-frequency load shedding (UFLS) for grid emergency control. We should recognize that DRL-based controls have notable advantage for most of the cases and time when it is well trained, but there will always be some extreme cases or corner cases that will be very difficult, if not impossible, for them to cover. Moving forward, we should leverage their complementary strengths. In this paper, we adopt a DRL- and rule-based hybrid grid emergency control approach where the existing rule-based controls are configured with conservative settings, e.g., a relatively longer time delay for action, to become a secondary or backup control, while DRL-based control acts as the primary control. We include these existing rule-based controls in the power system simulation environment for DRL training, such that DRL-based controllers can learn to coordinate with them.

\section{Test results}\label{sec:results}
Comprehensive tests are performed on a synthetic Texas power system with more than 3000 buses and over 50 GW total peak load. It is a modified system based on the original system developed in \cite{xu2017creation}. The main modifications include 1) adding sub-transmission and equivalent distribution low voltage buses and composite load models to better reflect the voltage stability issues in the real world; 2) adding hourly renewable generation output based on publicly available historical data. We created 1440 hourly power flow cases for 2 summer months. Details can be found in our earlier work \cite{sun2021sampling}.

\subsection{DRL-based emergency voltage control}
Our developed methods have been applied to different grid emergency controls including transient angular stability control via generator tripping, controlled islanding and emergency voltage control via adaptive load shedding. In this paper, our work is tested and demonstrated with the emergency voltage control via adaptive load shedding application. The goal is to develop a DRL-based closed-loop control policy for applying the load shedding at load centers including Houston areas, Dallas areas, Austin areas and San Antonio Areas to avoid the voltage stability issue and meet the voltage recovery requirements. 

The possible load shedding control actions are defined for a total of 258 buses with at least  50 MW dynamic motor loads in the following load centers: zone 3 (Houston, 81 buses), zone 16 (Dallas, 76 buses), at zone 23 (Austin, 61 buses), and zone 26 (San Antonio, 50 buses). The amount of load that could be shed for each bus at each action time step (0.1 s)  is a continuous variable from 0 (no load shedding) to 0.2 (shedding 20\% of the initial total load at the bus). As such, \textit{the dimension of the action space is 258}. The observations included voltage magnitudes at 115-kV and above buses in the four zones (total 468 buses) as well as the fractions of loads served at the 258 buses where load shedding could be applied. Thus,  \textit{the dimension of the observation space is 726}. 

The detailed mathematical problem formulation can be found in our earlier works\cite{huang2022accelerated,huang2022DMRL}. Compared to our earlier works, we adopt the same Markovian Decision Process (MDP) formulation, but we focus on developing and testing scalable methods based on the convergence framework for grid control at a large scale in this work.

\subsection{Dataset preparation}

 \begin{figure}[t!]
  \centering
    \includegraphics[width=0.25\textwidth]{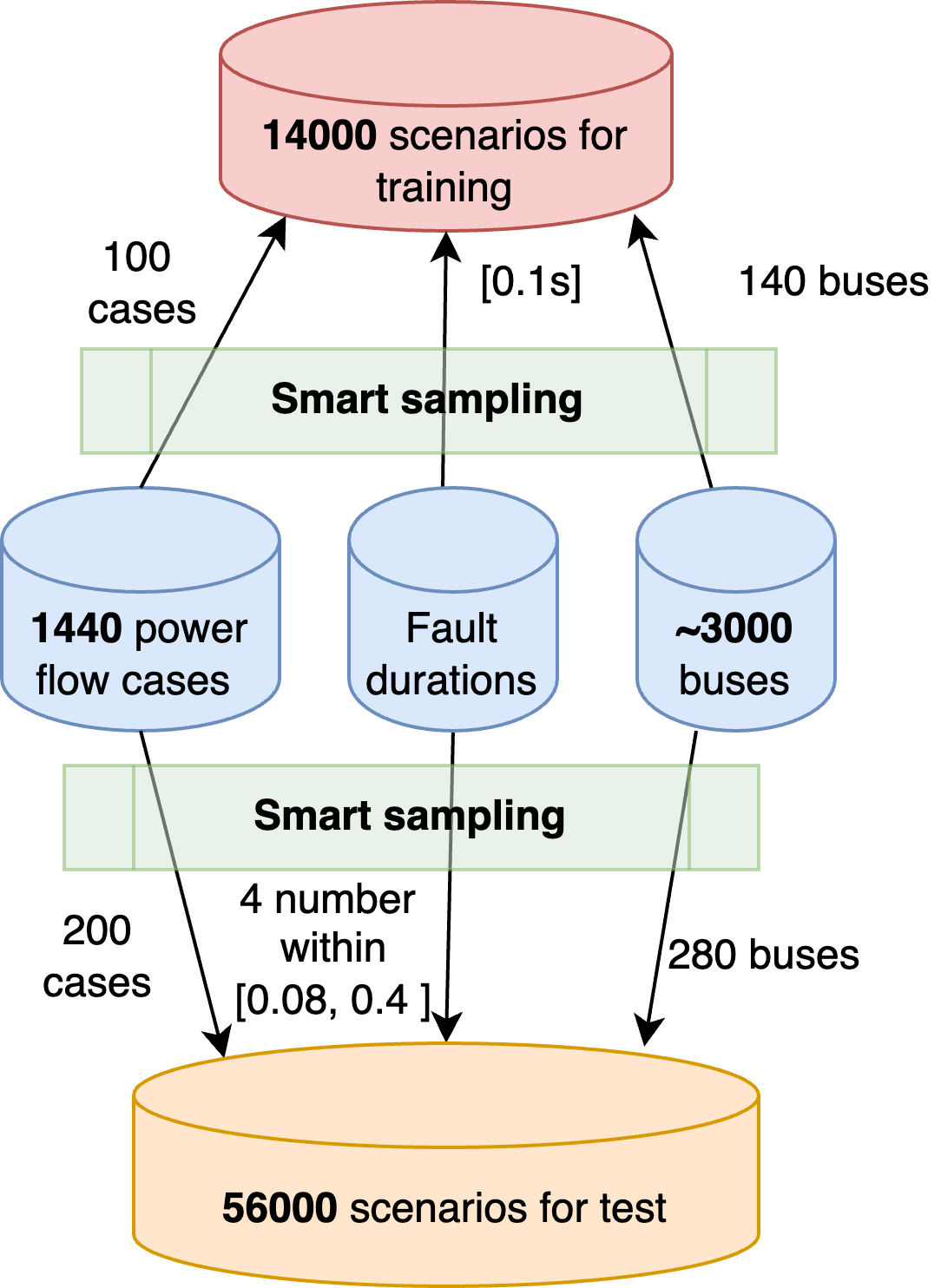}
  \caption{Preparation of the training and testing scenarios.}
  \label{fig:training_testing_scenarios}
 \end{figure}
 
 We employed the smart sampling method discussed in Section \ref{ssec:sampling} to create scenarios for training and testing, as shown in Fig. \ref{fig:training_testing_scenarios}. The training dataset includes 14000 different tasks (scenarios), which are combinations of 140 candidate fault buses and 100 different power flow cases. To comprehensively test the performance and generalization capability of the control policy, we sampled a larger dataset with 56000 scenarios that were unseen in the training.

\subsection{Training results}
\begin{figure}[!t]
  \centering
    \includegraphics[width=0.40\textwidth]{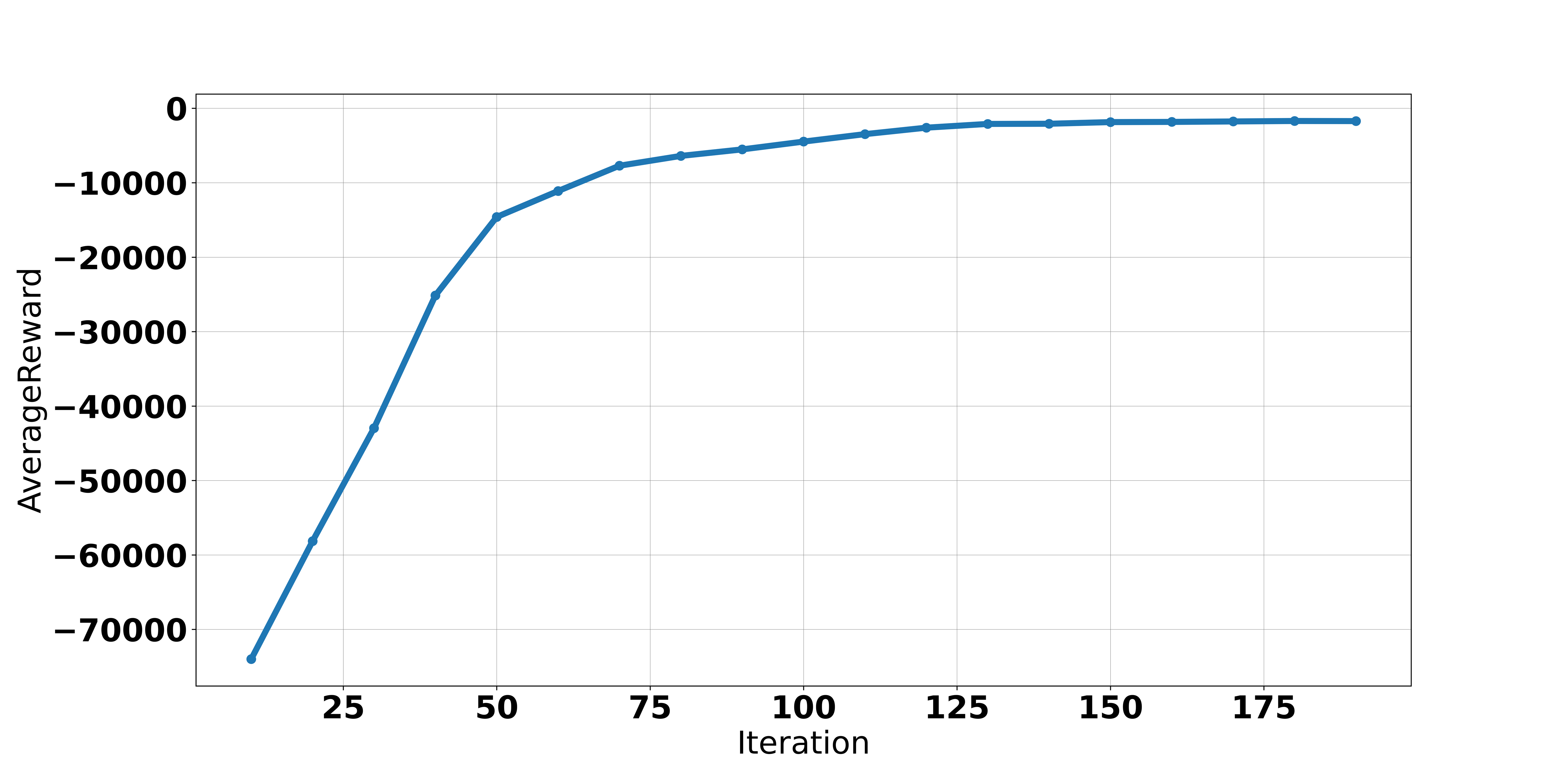}
  \caption{Training average reward over iterations.}
  \label{fig:Training_average_reward}
 \end{figure}

We performed the training and testing on an HPC cluster and used 840 CPU cores for training. At each of the training iterations, 40 fault locations are sampled from the 140 fault location candidates and 30 power flow cases are sampled from the 100 power flow candidates and the fault locations and power flow cases are combined to create the rollout tasks (1200 tasks per training iteration). Fig. \ref{fig:Training_average_reward} shows the learning curve of the training, which indicates the training converges around 200 iterations, and it takes about 110 hours to finish the training.


We also compared the meta-PARS-based load shedding control versus the existing rule-based UVLS load shedding scheme. We chose UVLS as the baseline as other existing optimal control methods such as model predictive control and DRL methods such as Proximal Policy Optimization (PPO) do not scale well to solve this complex problem using the available computing resources within a reasonable period.

\begin{figure}[!t]
  \centering
    \includegraphics[width=0.35\textwidth]{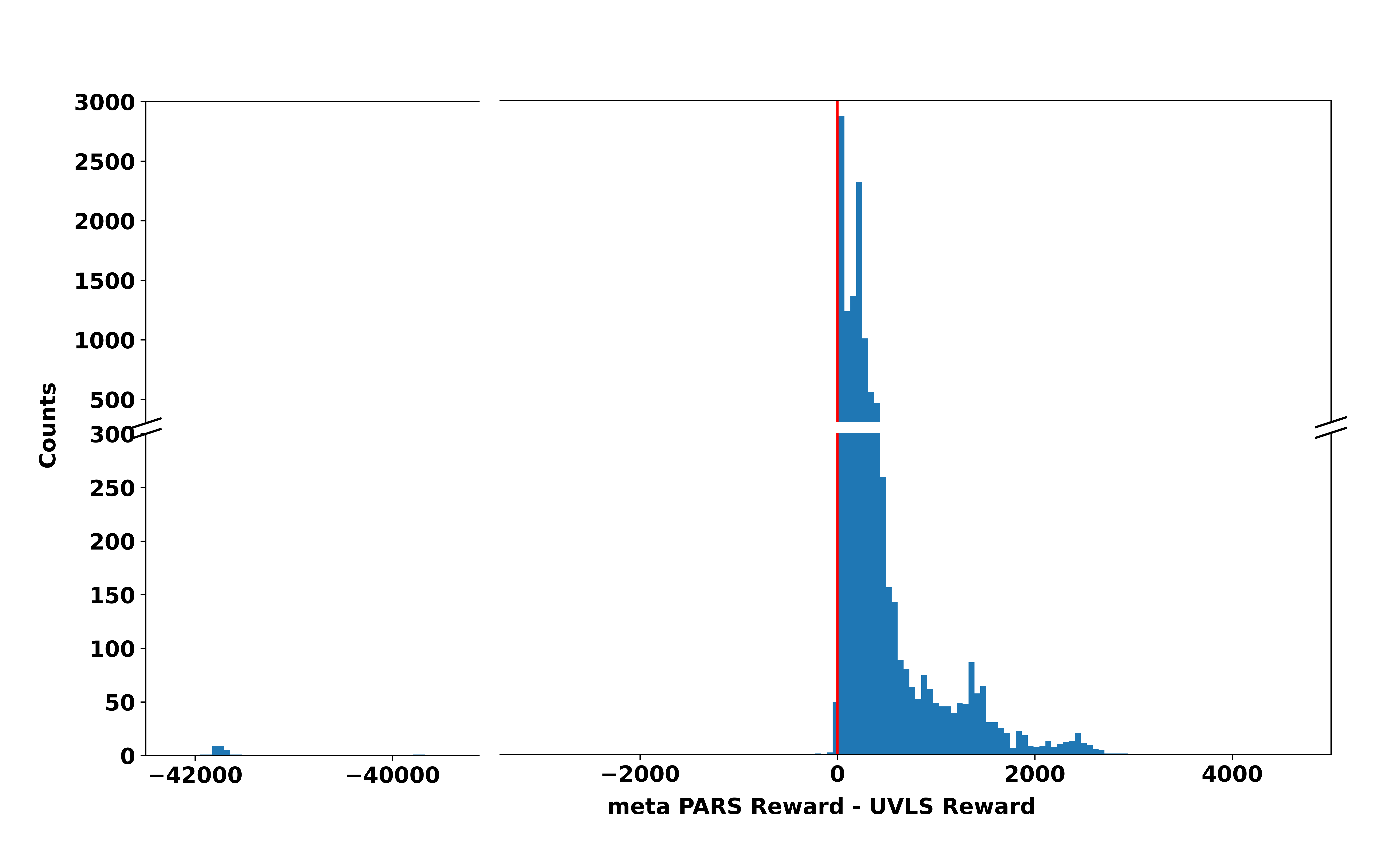}
  \caption{Histogram of reward differences between meta-PARS and UVLS methods for the training cases.}
  \label{fig:Histogram_meta_PARS_UVLS}
 \end{figure}

Among the 14000 training scenarios, there are 2687 scenarios do not require load shedding actions, and the meta-PARS does not provide actions for all the 2687 cases. For the 11313 scenarios do require load shedding actions to bring the voltage profile back to normal. To show the comparison results, we calculated the reward differences (i.e., the reward of meta-PARS subtracts that of UVLS) for all the total of 14000 trading tasks. A positive value means the our method is better for the corresponding test scenario and vice versa. Fig. \ref{fig:Histogram_meta_PARS_UVLS} shows the histogram of the rewards differences. Our method outperformed UVLS for (99.74\%) of the scenarios. 

\begin{figure}[!t]
  \centering
    \includegraphics[width=0.45\textwidth]{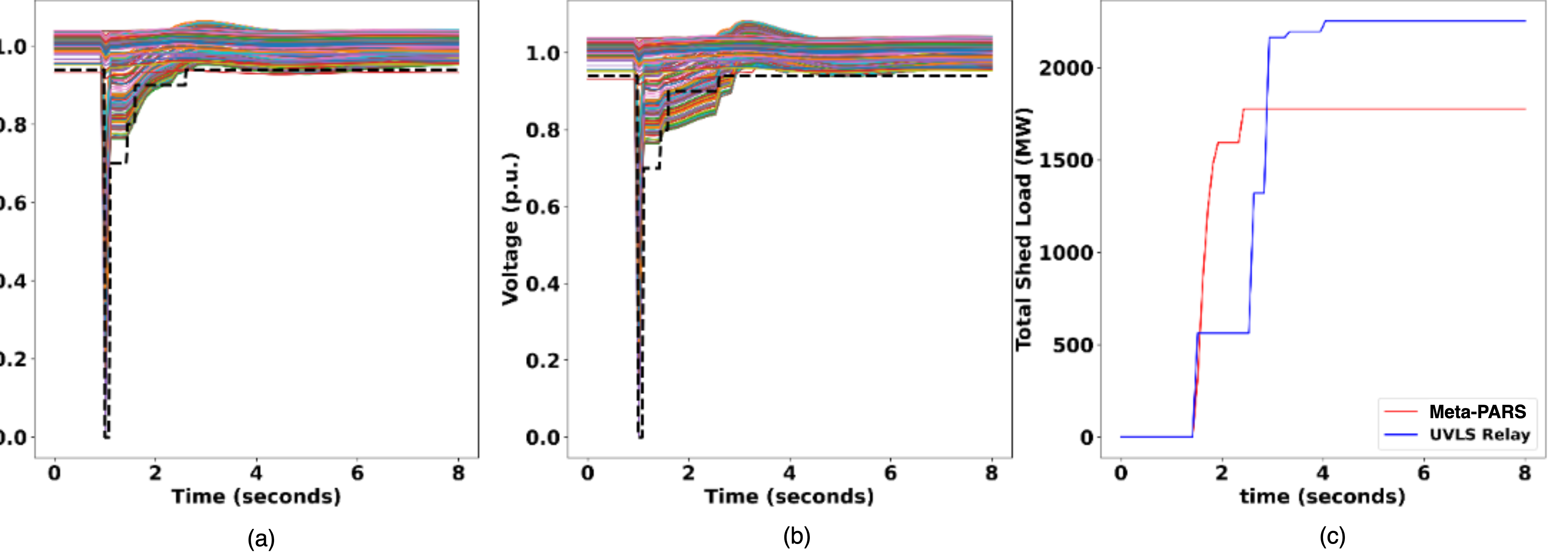}
  \caption{Comparison of meta-PARS and UVLS-based controls with power flow case 5 and a fault of duration 0.1s at bus 7037 in the training dataset: (a) bus voltages with meta-PARS-based control; (b) bus voltages with UVLS control; (c) total load shedding amount with meta-PARS and UVLS control methods.}
  \label{fig:training_voltage_recovery}
 \end{figure}

Fig. \ref{fig:training_voltage_recovery} shows the comparison between the meta-PARS and UVLS performance for a specific training task with a three-phase fault at 500 kV bus 7037. The total rewards of the meta-PARS and UVLS relay control in this test task are -1663 and -2142, respectively, and a larger reward is better. Fig. \ref{fig:training_voltage_recovery} (a) and (b) show that the voltages in the system with UVLS control recover much slower than the voltages in the system with meta-PARS control. Fig. \ref{fig:training_voltage_recovery}(c) shows that the meta-PARS-based control only shed around 1700 MW of load to bring the system voltage back to meet the standard, while the UVLS control shed more than 2200 MW load but still could not recover the system voltage to the level required by the standard. This key difference is the shedding action time instant and location. 

\subsection{Test results}
Once the policy is trained, we tested the trained control policy on a set of 56000 different scenarios with the combination of 280 fault buses that are different from the training fault buses and 200 power flow cases that are different from the training power flow cases. We also compared the meta-PARS-based control versus the conventional UVLS scheme. 

\begin{figure}[!t]
  \centering
    \includegraphics[width=0.40\textwidth]{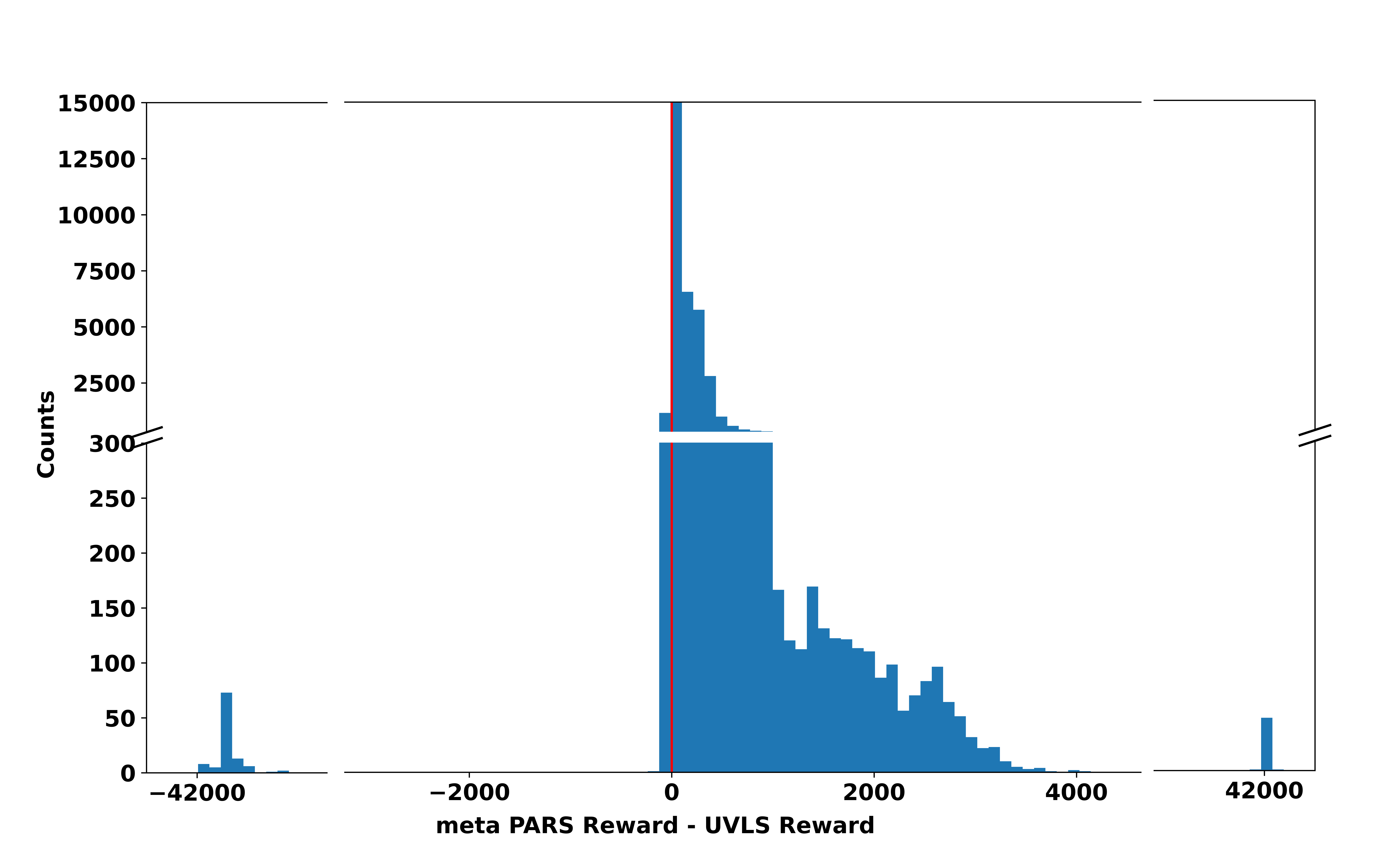}
  \caption{Histogram of the reward differences between meta-PARS and UVLS for the testing cases.}
  \label{fig:Histogram_meta_PARS_UVLS_testing}
 \end{figure}

Among the 56000 test scenarios, there are a total of 18962 scenarios that do not require load shedding actions for system recovery, and the meta-PARS does not provide load shedding actions for 18,914 of them (99.75\%). There are 37038 scenarios that require load shedding actions to meet the voltage recovery standard. In 94.147\% of them, the meta-PARS control outperforms the UVLS control. To show the comparison results, we calculated the reward differences (i.e., the reward of meta-PARS subtracts that of UVLS) for all the test tasks requiring load shedding actions. A positive value means the meta-PARS method is better for the corresponding test task and vice versa. Fig. \ref{fig:Histogram_meta_PARS_UVLS_testing} shows the histogram of the rewards differences. Our meta-PARS method outperforms UVLS scheme for most of the scenarios. At the same time, the result also shows that it is still necessary to have backup controls for a small number of extreme (or corner) cases. 

\begin{figure}[!t]
  \centering
    \includegraphics[width=0.45\textwidth]{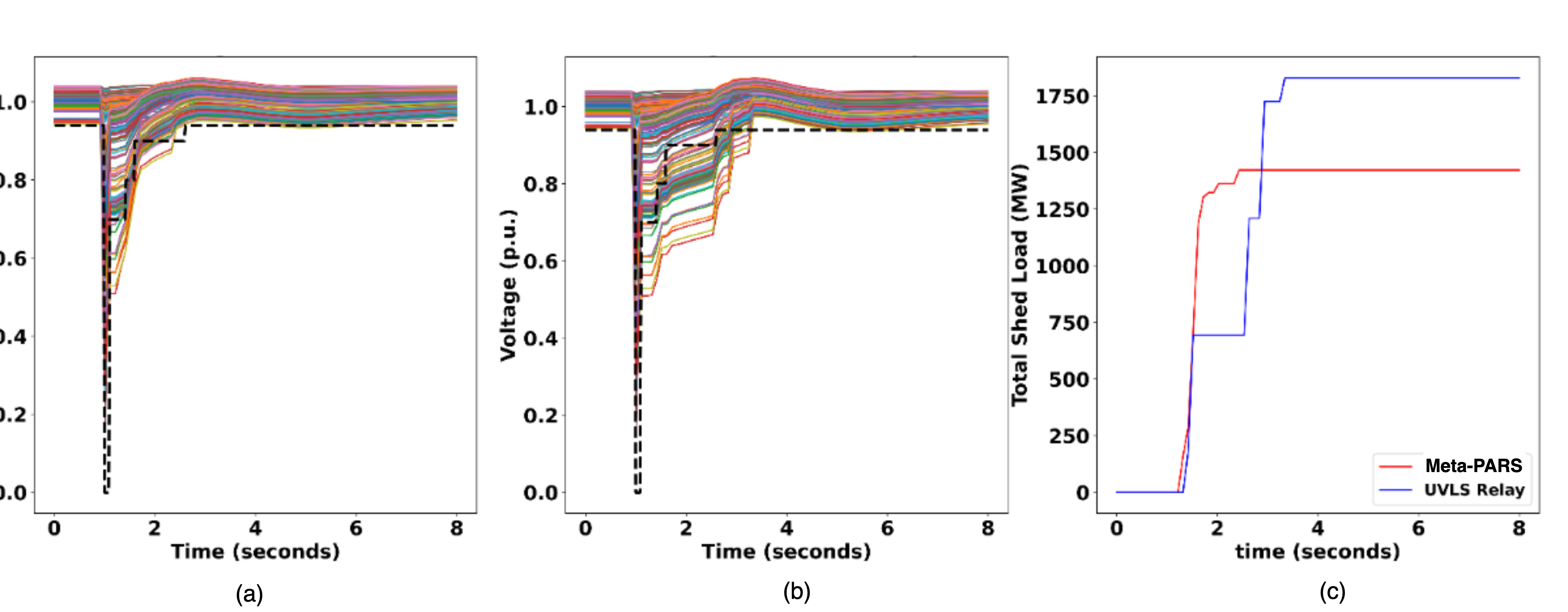}
  \caption{Comparison of meta-PARS and UVLS methods with the power flow case 9 and a 0.1 s fault at bus 5083 in the testing dataset: (a) bus voltages with meta-PARS-based control; (b) bus voltages with UVLS control; (c) total load shedding amount with meta-PARS and UVLS control methods.}
  \label{fig:testing_voltage_recovery}
 \end{figure}

Fig. \ref{fig:testing_voltage_recovery} shows the comparison of meta-PARS and UVLS performance for a specific test scenario with 0.1 s three-phase fault at the 500 kV bus 5083 in the Dallas region. The total rewards of the meta-PARS and UVLS relay control in this test task were -1439.5 and -1769.9, respectively. Fig. \ref{fig:testing_voltage_recovery} (a) and (b) show that the voltages in the system with UVLS control recover much slower compared with the voltages in the system with meta-PARS control. Fig. \ref{fig:testing_voltage_recovery}(c) shows that meta-PARS control only shed about 1300 MW loads to bring the system voltage back to meet the performance requirements, while the UVLS control shed more than 1750 MW loads but still cannot recover voltages to the required level. 


\section{Conclusions}
The rapid evolution of the grid and the push towards decarbonization necessitate advanced solutions to ensure the stability, reliability, and resilience of electricity services. This paper has delved into the pressing need for intelligent emergency control in large-scale power systems. We considered the DRL-based methods as the backbone for achieving intelligent grid control. However, there are multifaceted challenges in DRL-based control such as scalability, adaptiveness, and security posed by the complex power system landscape. To this end, the paper proposes and instantiates a convergence framework integrating power systems physics, machine learning, advanced computing, and grid control to realize intelligent grid control at a large scale. We instantiated the framework by developing several key, interdisciplinary methods and tools including scalable grid simulation environment, highly scalable meta-PARS method and physics-informed three-stage DRL agent training process.

The proposed solutions were rigorously tested on a synthetic Texas power system with more than 3000 buses, considering 14000 scenarios for training and 56000 unseen scenarios for testing. The results were promising, showing a 26\% reduction in load shedding on average compared to the existing rule-based UVLS control scheme, and outperforming the baseline in 99.7\% of the test scenarios in terms of the control objective value (total reward). The comprehensive methods and tools developed have demonstrated significant progress and have the potential to be a cornerstone in the evolution of intelligent grid control design. Future research directions include 1) adaptation and extension for other grid control applications and 2) collaboration between human operators and AI agents for grid operation and control.

\bibliographystyle{IEEEtran}
\bibliography{references.bib}

\begin{thebibliography}{10}
\providecommand{\url}[1]{#1}
\csname url@samestyle\endcsname
\providecommand{\newblock}{\relax}
\providecommand{\bibinfo}[2]{#2}
\providecommand{\BIBentrySTDinterwordspacing}{\spaceskip=0pt\relax}
\providecommand{\BIBentryALTinterwordstretchfactor}{4}
\providecommand{\BIBentryALTinterwordspacing}{\spaceskip=\fontdimen2\font plus
\BIBentryALTinterwordstretchfactor\fontdimen3\font minus \fontdimen4\font\relax}
\providecommand{\BIBforeignlanguage}[2]{{%
\expandafter\ifx\csname l@#1\endcsname\relax
\typeout{** WARNING: IEEEtran.bst: No hyphenation pattern has been}%
\typeout{** loaded for the language `#1'. Using the pattern for}%
\typeout{** the default language instead.}%
\else
\language=\csname l@#1\endcsname
\fi
#2}}
\providecommand{\BIBdecl}{\relax}
\BIBdecl

\bibitem{Yan_Frequency_DRL}
Z.~{Yan} and Y.~{Xu}, ``Data-driven load frequency control for stochastic power systems: A deep reinforcement learning method with continuous action search,'' \emph{IEEE Transactions on Power Systems}, vol.~34, no.~2, pp. 1653--1656, 2019.

\bibitem{huang2019loadshedding_DRL}
Q.~{Huang}, R.~{Huang}, W.~{Hao}, J.~{Tan}, R.~{Fan}, and Z.~{Huang}, ``Adaptive power system emergency control using deep reinforcement learning,'' \emph{IEEE Transactions on Smart Grid}, vol.~11, no.~2, pp. 1171--1182, 2020.

\bibitem{chen2020model}
C.~Chen, M.~Cui, F.~F. Li, S.~Yin, and X.~Wang, ``Model-free emergency frequency control based on reinforcement learning,'' \emph{IEEE Transactions on Industrial Informatics}, 2020.

\bibitem{glavic2019DRL4GC}
M.~Glavic, ``(deep) reinforcement learning for electric power system control and related problems: A short review and perspectives,'' \emph{Annual Reviews in Control}, 2019.

\bibitem{Chen2022RL}
X.~Chen, G.~Qu, Y.~Tang, S.~Low, and N.~Li, ``Reinforcement learning for selective key applications in power systems: Recent advances and future challenges,'' \emph{IEEE Transactions on Smart Grid}, vol.~13, no.~4, pp. 2935--2958, 2022.

\bibitem{li2023DRL}
Y.~Li, C.~Yu, M.~Shahidehpour, T.~Yang, Z.~Zeng, and T.~Chai, ``Deep reinforcement learning for smart grid operations: Algorithms, applications, and prospects,'' \emph{Proceedings of the IEEE}, vol. 111, no.~9, pp. 1055--1096, 2023.

\bibitem{huang2022accelerated}
R.~Huang, Y.~Chen, T.~Yin, X.~Li, A.~Li, J.~Tan, W.~Yu, Y.~Liu, and Q.~Huang, ``Accelerated derivative-free deep reinforcement learning for large-scale grid emergency voltage control,'' \emph{IEEE Transactions on Power Systems}, vol.~37, no.~1, pp. 14--25, 2022.

\bibitem{weng2022envpool}
J.~Weng, M.~Lin, S.~Huang, B.~Liu, D.~Makoviichuk, V.~Makoviychuk, Z.~Liu, Y.~Song, T.~Luo, Y.~Jiang, Z.~Xu, and S.~Yan, ``Envpool: A highly parallel reinforcement learning environment execution engine,'' 2022.

\bibitem{Ray}
Anyscale, ``{Ray}: An open-source unified framework for scaling ai and python applications,'' \url{https://www.ray.io/}, accessed: 2023-9-27.

\bibitem{GridPACK}
B.~Palmer, W.~Perkins, Y.~Chen, S.~Jin, D.~Callahan, K.~Glass, R.~Diao, M.~Rice, S.~Elbert, M.~Vallem, and Z.~Huang, ``Gridpack: A framework for developing power grid simulations on high performance computing platforms,'' in \emph{2014 Fourth International Workshop on Domain-Specific Languages and High-Level Frameworks for High Performance Computing}, 2014, pp. 68--77.

\bibitem{GridPACK-source}
\BIBentryALTinterwordspacing
{PNNL}, ``{GridPACK}.'' [Online]. Available: \url{https://github.com/GridOPTICS/GridPACK}
\BIBentrySTDinterwordspacing

\bibitem{huang2022DMRL}
R.~Huang, Y.~Chen, T.~Yin, Q.~Huang, J.~Tan, W.~Yu, X.~Li, A.~Li, and Y.~Du, ``Learning and fast adaptation for grid emergency control via deep meta reinforcement learning,'' \emph{IEEE Transactions on Power Systems}, vol.~37, no.~6, pp. 4168--4178, 2022.

\bibitem{HADREC-source}
\BIBentryALTinterwordspacing
{Huang, Renke and Huang, Qiuhua and Yin, Tianzhixi, and Palmer, Bruce, and Li, Ang}, ``{GridPACK}.'' [Online]. Available: \url{https://github.com/pnnl/hadrec}
\BIBentrySTDinterwordspacing

\bibitem{WECC_voltage}
{WECC}, ``Tpl-001-wecc-crt-4—transmission system planning performance,'' \url{https://www.wecc.org/Reliability/TPL-001-WECC-CRT-4.pdf}, accessed: 2023-9-28.

\bibitem{du2022PIES}
Y.~Du, Q.~Huang, R.~Huang, T.~Yin, J.~Tan, W.~Yu, and X.~Li, ``Physics-informed evolutionary strategy based control for mitigating delayed voltage recovery,'' \emph{IEEE Transactions on Power Systems}, vol.~37, no.~5, pp. 3516--3527, 2022.

\bibitem{xu2017creation}
T.~Xu, A.~B. Birchfield, K.~S. Shetye, and T.~J. Overbye, ``Creation of synthetic electric grid models for transient stability studies,'' in \emph{The 10th Bulk Power Systems Dynamics and Control Symposium (IREP 2017)}, 2017, pp. 1--6.

\bibitem{wang2021curriculum}
X.~Wang, Y.~Chen, and W.~Zhu, ``A survey on curriculum learning,'' \emph{IEEE Transactions on Pattern Analysis and Machine Intelligence}, vol.~44, no.~9, pp. 4555--4576, 2021.

\bibitem{sun2021sampling}
X.~Sun, X.~Li, S.~Datta, X.~Ke, Q.~Huang, R.~Huang, and Z.~J. Hou, ``Smart sampling for reduced and representative power system scenario selection,'' \emph{IEEE Open Access Journal of Power and Energy}, vol.~8, pp. 293--302, 2021.

\end{thebibliography}

\end{document}